\documentclass{article}
\usepackage{graphicx, enumitem} 

\usepackage{amsthm,amsmath,amsfonts,amssymb}

\usepackage{pgfplots}
\pgfplotsset{compat=1.18}

\usepackage{hyperref,booktabs}
\usepackage{fullpage}
\usepackage{algorithm}
\usepackage{algorithmic}

\newtheorem{theorem}{Theorem}

\title{How Many Submissions May an Author Make?
\\A Harmonic Quota for Submissions under Coauthorship}
\author{Nihar B. Shah\\Carnegie Mellon University\\nihars@cs.cmu.edu}
\date{}

\begin{document}

\maketitle

\begin{abstract}
Research evaluation systems---including journals, conferences, and funders---are increasingly using author-level submission limits to manage growing submission loads. Most existing policies charge each submission as
a unit cost against every coauthor’s quota. This treats a solo-authored submission and a large collaborative submission identically for each author, even though the reviewing demand of a collaborative submission is jointly attributable to many authors rather than one. Thus we ask the question: how many submissions may an author make under coauthorships? We propose a 
\emph{Harmonic Quota Rule}, in which an author's cost for a
submission decreases with the number of coauthors as the reciprocal of their
harmonic number. We derive this rule in a principled manner that navigates the tension between respecting collaborations and being resistant to manipulation by adding spurious authors. We also develop
a Generalized Harmonic Quota Rule, a framework that subsumes the Harmonic Quota Rule and other natural quota rules. Our framework
requires specification of only three interpretable parameters, thereby enabling organizers to choose among various seemingly disparate rules. Our work may also be useful in other scarce-resource allocation settings, such as allocation of compute and telescope time.\\
\begin{center}{\tt Interactive tool:} \url{https://www.cs.cmu.edu/~nihars/quota/organizer.html}
\end{center}
\end{abstract}

\section{Introduction}

Scientific journals, conferences, and funding agencies have recently seen a significant surge in submissions~\cite{shah2025survey}. Due to the increased workload on evaluators, these bodies (who we will collectively refer to as ``organizers'') are imposing various restrictions on submissions. A prominent approach is to impose fixed quotas on authors: each author can make at most a fixed number of submissions~\cite{nih2025cap,nsf2024cise,cvpr2026author,iccv2025author,pvldb2027guidelines,comsoc_twc_policies,ieeesp2027cfp}. For instance, the VLDB conference is allowing authors to submit at most twelve papers per year, while the National Institutes of Health is allowing researchers to submit at most 6 grant applications per year. 

We refer to this popular approach as the \emph{fixed quota rule} where each author is allowed at most a certain number of submissions. While the fixed quota rule is simple, it is problematic because it ignores effects of coauthorships. It treats a solo-authored submission and a large collaborative submission identically for each author, even though each author in a large collaborative submission should bear only a share of the reviewing demand, and typically receives diluted credit for the submission, as compared to a solo author. If a senior researcher who supervises many projects reaches the cap, junior collaborators may be blocked from submitting work even when they have not overused the system~\cite{cao2025dissecting}. A quota rule should recognize these distinctions.
 
Another natural baseline that we will use for comparison is what we call the \emph{per-capita quota rule}. This rule treats each submission as a unit cost that is distributed across authors. More precisely, a submission with $a$ authors contributes $\frac{1}{a}$ to each authors' quota usage. Authors may continue to submit until they exhaust a pre-defined maximum quota. The problem with this per-capita rule is that it is highly vulnerable to gaming. Since the cost falls in direct proportion to the number of authors, an author's submission allowance grows rapidly -- linearly with every added coauthor. Padding author lists with spurious coauthors would then let an unscrupulous author accumulate much more total credit overall.\\

With this context, our \textbf{main contributions} are as follows. 
\begin{itemize}
    \item We formulate the problem of designing coauthorship-sensitive submission quota rules, where the quota cost of a submission may depend on the number of coauthors. We approach the design of such rules by balancing (i) respecting collaborations -- a solo-authored submission should not cost the same as large collaboration for an author, with (ii) resistance to manipulation -- an author cannot increase total credit they can earn by adding spurious coauthors.
    
    \item We propose a Harmonic Quota Rule for the number of submissions any author can make. In this rule, the cost of a submission decreases harmonically in the number of authors. Our rule is derived in a principled manner based on an empirically-validated model of credit assignment among coauthors.
    \item We then develop a Generalized Harmonic Quota Rule, a more general framework that encompasses various rules such as the fixed, per-capita, and harmonic. Even though these individual rules appear quite disparate, our general framework unifies them and allows for choosing between these and other rules by simply specifying three parameters that are easy to reason about. 

    \item Finally, we provide an interactive tool for using these quota rules. Organizers can use this tool to specify a rule and its parameters, and can generate a link for authors to check their submissions against the specified rule. The tool is available at \url{https://www.cs.cmu.edu/~nihars/quota/organizer.html}.
\end{itemize}

This quota-design problem also arises beyond journals, conferences, and funding programs. Many other systems allocate scarce resources through proposals submitted by individuals or teams, often followed by peer review or committee evaluation. Examples include institutional compute clusters allocated to research groups~\cite{Acc26,INC26}, telescope time allocated to astronomers through reviewed proposals~\cite{carpenter2025enhancing}, and others~\cite{NSFShipTime24,ALSGeneralUser26,NSLSIIUserAccess26}. We envisage that our rules may also be useful in some such settings, although their deployment would need to be adapted to the norms and constraints of each domain.

We present our proposed Harmonic Quota Rule and Generalized Harmonic Quota framework in Section~\ref{Sec:Rule}. We present the derivation of our Harmonic Quota Rule in Section~\ref{Sec:Derivation} and that of our Generalized Harmonic Quota framework in Section~\ref{Sec:General_derivation}. We conclude with a discussion and pointers towards future work in Section~\ref{Sec:Practical}.

\section{Proposed rules} \label{Sec:Rule}
In this section, we specify our Harmonic Quota Rule, followed by our generalization, and then present a plot illustrating the various rules. 

\subsection{Harmonic Quota Rule}

Our proposed rule is enumerated as Algorithm~\ref{alg:harmonic_quota}. It takes two parameters as inputs from the organizers, representing two extremes of coauthorship. The first parameter is the number of submissions that any author can make if they only submit single-author papers. The second parameter, $N_\infty$, is an absolute ceiling on submissions---the
number an author could make in the limit of infinitely many coauthors, and a
bound no author can exceed however many coauthors they add.

\begin{algorithm}[H]
\caption{Harmonic Quota Rule}\label{alg:harmonic_quota}
\begin{algorithmic}[1]
 
\REQUIRE Two parameters set by the peer-review organizers:
\begin{itemize}[noitemsep,topsep=0pt] 
\itemindent=-13pt
\item $N_1$: maximum submissions for author submitting only single-authored submissions
\item $N_{\infty}$: maximum submissions for an author appearing with infinitely many coauthors~~ ($N_{\infty} \geq N_1$)
 \end{itemize}
 
\medskip
\STATE \textbf{Quota.} Each author begins with a budget of $N_1$.
 
\medskip
\STATE \textbf{Submission cost.} A submission with $a$ authors costs each of its authors
\[
f(a) \;=\; \frac{N_1}{N_{\infty}} \;+\; \left(1 - \frac{N_1}{N_{\infty}}\right)\frac{1}{H_a},
\qquad \text{where~} H_a = \sum_{j=1}^{a}\frac{1}{j}.
\]
 
\medskip
\STATE \textbf{Submission rule.} A submission is permitted if and only if each author has a budget  of at least $f(a)$ remaining.  Upon submission, $f(a)$ is deducted from each author's budget.
 
\end{algorithmic}
\end{algorithm}
 
Let us consider an example. Suppose one allows fully solo authors to make $2$ submissions, whereas authors in collaborations of infinitely many authors are allowed to make $20$ submissions. Then the organizers will set $N_1 = 2$ and $N_{\infty} = 20$. These are the only two values that the organizers need to choose. Now under our Harmonic rule, each author will initially have a budget of $N_1 = 2$. A submission with $a$ authors will cost every author $f(a) = 0.1 + \frac{0.9}{H_a}$. For instance, if the submission has $a= 5$ authors, then $H_a = \frac{137}{60}$, and it will cost each author $\frac{677}{1370} \approx 0.49$ out of their budget of $2$. 

\subsection{A generalized framework and rule} 
In this section, we generalize our Harmonic Quota Rule from Algorithm~\ref{alg:harmonic_quota}. Our generalization stems from two key motivations.

First, recall that in Algorithm~\ref{alg:harmonic_quota}, the parameter $N_{\infty}$ represents the total number of submissions that a researcher is permitted to make if all their submissions have infinite coauthors. While this hypothetical setting may be easy to envisage sometimes, organizers may alternatively want to consider more realistic settings. In particular, consider submissions with a certain realistic collaboration size, where this size to be thought about is determined by the organizers. Then in our setting, instead of imagining infinite authors, the organizers need to specify the number of submissions an author in the chosen collaboration size can make. 

Second, there are other rules such as the fixed and per-capita rules along with the Harmonic Quota Rule. All of these rules seem quite disparate. We thus wish to create a general and easy-to-use framework that can allow the organizers to choose between these or other rules. 

We thus generalize our Harmonic Quota Rule where in addition to the number of submissions $N_1$ that can be made by solo authors, the input requires organizers to specify some value $A$ of their choice, and another value $N_A$ as the number of submissions a researcher can make if all their submissions have $A$ authors.

We present this rule in Algorithm~\ref{alg:generalized_harmonic_quota}. As we discuss below, this rule subsumes Algorithm~\ref{alg:harmonic_quota} as well as other intuitive rules. Our framework thus allows to choose between all of these rules based on specifying only the values of $N_1$, $A$, and $N_A$, which are easy for organizers to reason about. 

\begin{algorithm}[H]
\caption{Generalized Harmonic Submission Quota}\label{alg:generalized_harmonic_quota}
\begin{algorithmic}[1]
 
\REQUIRE Three parameters set by the peer-review organizers:
\begin{itemize}[noitemsep,topsep=0pt] 
\itemindent=-13pt
\item $N_1$: maximum submissions for author making only single-authored submissions
\item $A$: number of authors in ``large'' collaborations~~ ($A \geq 2$)
\item $N_A$: maximum submissions for author whose submission each have $A$ authors~~($N_1 \leq N_A \leq A N_1$)
 \end{itemize}
\medskip
\STATE \textbf{Pre-processing.} Let $H_{a,p} = \sum_{j=1}^{a}\frac{1}{j^p}$ denote the $a^{th}$ generalized harmonic number of order $p$.
\IF{$H_{A,1} > \frac{N_A}{N_1}$}
\STATE Set $p=1$ and $N_{\infty} =N_A \;  \frac{1 - \frac{1}{H_{A,1}}}{1 - \frac{N_A}{N_1}\frac{1}{H_{A,1}}}$.
\ELSE
\STATE Set $p$ to be the unique value such that $H_{A,p} = \frac{N_A}{N_1}$, and set $N_{\infty} = \infty$.
\ENDIF
\medskip
\STATE \textbf{Quota.} Each author begins with a budget of $N_1$.
\medskip
\STATE \textbf{Submission cost.} A submission with $a$ authors costs each of its authors
\[
f(a) \;=\; \frac{N_1}{N_{\infty}} \;+\; \left(1 - \frac{N_1}{N_{\infty}}\right)\frac{1}{H_{a,p}}.
\]
\medskip
\STATE \textbf{Submission rule.} A submission is permitted if and only if each author has a budget of at least $f(a)$ remaining.  Upon submission, $f(a)$ is deducted from each author's budget.
\end{algorithmic}
\end{algorithm}
The value of $p$ in the else condition can be approximated easily, for instance, via binary search since $H_{A,p}$ strictly decreases with an increase in the value of $p$. Note that equivalently, the chosen $p$ is the largest value such that $H_{A,p}\geq \frac{N_A}{N_1}$. As the example below shows, one may have to make some approximations, and this can be done by pre-specifying a tolerance factor, e.g., flooring after the third decimal place. 

We note that the generality of this framework comes with a tradeoff. While we use resistance to manipulation as a design principle for Algorithm~\ref{alg:harmonic_quota}, our generalized framework may not always satisfy it. It holds only when $p=1$. When $p<1$, the per-author cost decays faster
than $1/H_a$, and padding author lists can increase an author's total
accumulated credit. The parameter $N_A$ thus situates the rule between
the manipulation-resistant harmonic rule and the gameable per-capita rule, 
letting organizers navigate the tension between respecting collaborations and
resisting manipulation.

Let us now consider an example. Suppose the organizers specify that solo authors can make at most $N_1=2$ submissions, and that collaborations involving $A=20$ authors each can make at most $N_A=10$ submissions. In this case, we have $H_{A,1} \approx 3.6$ and $N_1 H_{A,1} \leq N_A$. Consequently, the algorithm sets $p=0.759$ and $N_\infty = \infty$. Now, each author begins with a budget of $N_1=2$. A submission with $a$ authors costs each author $f(a) = \frac{1}{H_{a,0.759}}$. For instance, if $a=5$, then it incurs a cost of approximately $0.374$ to each author.

~\\We now discuss special cases of our Generalized Harmonic Submission Quota (Algorithm~\ref{alg:generalized_harmonic_quota}): 
\begin{itemize}
    \item Harmonic Quota Rule (Algorithm~\ref{alg:harmonic_quota}):  Recovered as the limiting case
$A=\infty$, with $N_A$ then interpreted as $N_\infty$.

    \item Per-capita quota rule (submission with $a$ authors counts as $\frac{1}{a}$ for each author): Obtained when $N_A = A N_1$, which leads to $p=0$ and hence $f(a) = \frac{1}{a}$.
    \item Fixed quota rule (each submission counts as $1$ for each author): Obtained when $N_A = N_1$, which results in $N_{\infty}= N_1$ and hence $f(a) = 1$. 
\end{itemize}

\subsection{A plot} 

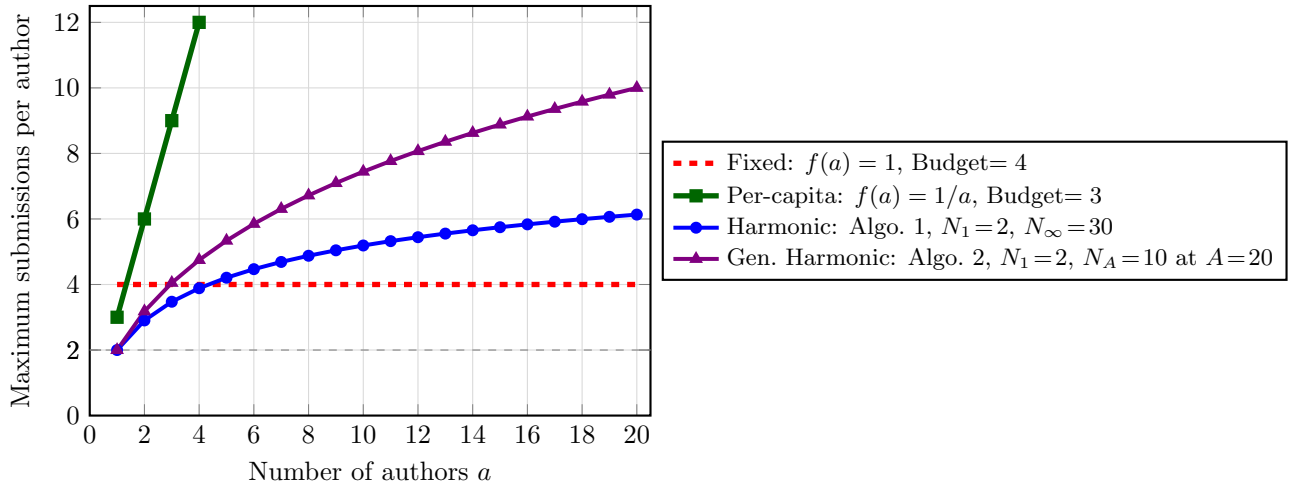
\begin{figure}[h]
    \centering
\begin{tikzpicture}
\begin{axis}[
    xlabel={Number of authors $a$},
    ylabel={Maximum submissions per author},
    xmin=0, xmax=20.5,
    ymin=0, ymax=12.5,
    ytick={0,2,4,6,8,10,12},
    grid=major,
    grid style={gray!30},
    width=9cm,
    height=7cm,
    thick,
    extra y ticks={2},
    extra y tick style={
        grid=major,
        grid style={dashed, black!50},
        tick label style={anchor=east},
    },
    legend style={at={(1.02,0.5)}, anchor=west, font=\small, cells={anchor=west}},
]

\addplot[red, dashed, line width=2pt] coordinates {
    (1, 4.0000)
    (20, 4.0000)
};
\addlegendentry{Fixed: $f(a)=1$, Budget$=4$}

\addplot[black!60!green, line width=2pt, mark=square*, mark size=1.5pt] coordinates {
    (1, 3.0000)
    (2, 6.0000)
    (3, 9.0000)
    (4, 12.0000)
};
\addlegendentry{Per-capita: $f(a)=1/a$, Budget$=3$}

\addplot[blue, line width=1.5pt, mark=*, mark size=1.5pt] coordinates {
    (1, 2.0000)
    (2, 2.9032)
    (3, 3.4737)
    (4, 3.8860)
    (5, 4.2068)
    (6, 4.4681)
    (7, 4.6879)
    (8, 4.8772)
    (9, 5.0430)
    (10, 5.1905)
    (11, 5.3230)
    (12, 5.4432)
    (13, 5.5532)
    (14, 5.6544)
    (15, 5.7481)
    (16, 5.8353)
    (17, 5.9168)
    (18, 5.9933)
    (19, 6.0653)
    (20, 6.1333)
};
\addlegendentry{Harmonic: Algo.~\ref{alg:harmonic_quota}, $N_1\!=\!2$, $N_{\infty}\!=\!30$}

\addplot[violet, line width=1.5pt, mark=triangle*, mark size=1.5pt] coordinates {
    (1, 2.0000)
    (2, 3.1815)
    (3, 4.0499)
    (4, 4.7479)
    (5, 5.3372)
    (6, 5.8502)
    (7, 6.3065)
    (8, 6.7189)
    (9, 7.0960)
    (10, 7.4441)
    (11, 7.7678)
    (12, 8.0709)
    (13, 8.3561)
    (14, 8.6257)
    (15, 8.8816)
    (16, 9.1252)
    (17, 9.3578)
    (18, 9.5806)
    (19, 9.7944)
    (20, 10.0000)
};
\addlegendentry{Gen.\ Harmonic: Algo.~\ref{alg:generalized_harmonic_quota}, $N_1\!=\!2$, $N_A\!=\!10$ at $A\!=\!20$}
\end{axis}
\end{tikzpicture}
    \caption{Maximum number of submissions an author can make if each of their submissions has $a$ authors. }
    \label{fig:comparison}
\end{figure}

Figure~\ref{fig:comparison} illustrates these rules by plotting the number of submissions that an author can make for a fixed collaboration size.  Observe that the fixed quota is flat since it treats collaborative and solo authorship identically. Per-capita counting grows linearly. The harmonic quota grows logarithmically. The generalized harmonic quota lies between the harmonic and per-capita cases, reaching exactly $10$ submissions at $a=20$.

This plot also shows another reason why one may wish to use the generalized harmonic framework. In the harmonic quota rule, the number of submissions per author grows only logarithmically in the number of authors, which may be considered too slow by some organizers. The generalized harmonic framework enables organizers to choose rules that grow faster. 

\section{Derivation of Harmonic Quota Rule (Algorithm~\ref{alg:harmonic_quota})} \label{Sec:Derivation}

In this section, we present our derivation of Algorithm~\ref{alg:harmonic_quota}.

\subsection{Notation} 
We let $k$ denote the initial budget for each author. In the context of rolling submissions like in a journal, this budget can be applied in a calendar year or in a windowed fashion. If a submission has $a$ authors, then the submission incurs a cost of $f(a)$ from the budget of each author, for some function $f:\mathbb{Z}_{>0} \rightarrow \mathbb{R}_{\geq 0}$. In order to make a submission, each author of that submission should have a non-negative quota after counting the current submission. Our goal is to derive the function $f$, in conjunction with the budget $k$, in a principled manner.

Note that the fixed quota rule that is deployed in conferences and funding calls today falls under this class with $f(a) =1$ and $k$ set as the specified budget. Another comparison point is the per-capita quota rule where each submission incurs a unit cost that is distributed evenly across all authors, yielding $f(a) = \frac{1}{a}$.

\subsection{Resistance to manipulation: Adding spurious authors should not increase achievable credit}\label{SecStrategyproof}
For a given quota rule, an unscrupulous author could add more coauthors in order to reduce the contribution of each submission to their quota and increase the number of submissions they can make. If the quota used decreases too fast with increase in authorship size, such authors can earn more cumulative credit  by gaming the rule in this manner. We wish to design the quota rule $f$ that is resistant to such manipulation.

We call upon~\cite{hodge1981publication, hagen2008harmonic, hagen2010harmonic, hagen2013harmonic} for an empirically verified model of credit among coauthors. This model was first proposed by Hodge and Greenberg~\cite{hodge1981publication}, and subsequently validated by Hagen against survey data 
on perceived authorship credit across several fields~\cite{hagen2008harmonic, hagen2010harmonic, hagen2013harmonic}. This model assigns the lead author of a 
submission with $a$ authors a credit share of $\frac{1}{H_a}$, where $H_a=\sum_{j=1}^{a}\frac1j$ is the $a^{th}$ harmonic number. More generally, the $i^{th}$ ranked author gets a credit share $\frac{1/i}{H_a}$. 

Now consider an author who makes single author submissions. They can make $\frac{k}{f(1)}$ submissions. They can increase their quota by simply adding more authors to their submissions. Suppose they add $a-1$ authors to each submission, then they can now make $\frac{k}{f(a)}$ submissions. However, the addition of new authors dilutes their perceived contribution to, and hence credit for, each submission. The total credit they can accumulate by making single-author submissions is $\frac{k}{f(1)} \times \frac{1}{H_1}$. By adding $a-1$ more authors, and assuming they continue to be the lead author, the total credit they can accumulate is $\frac{k}{f(a)} \times \frac{1}{H_a}$. 

The inclusion of these additional authors may allow the original author to make more submissions, but should not increase the total credit for them. Thus we need:
\begin{align*}
    \frac{k}{f(1)} \times \frac{1}{H_1} \geq \frac{k}{f(a)} \times \frac{1}{H_a},
\end{align*}
and using the fact that $H_1=1$, we get
\begin{align}
    f(a) \geq \frac{f(1)}{H_a}.
    \label{Eqnfundamental}
\end{align}
This inequality is a necessary condition for resistance to manipulation: 
any cost that decays faster than $\frac{f(1)}{H_a}$ as coauthors are added would let an
author inflate their total accumulated credit by padding author lists. In contrast, for instance under the per-capita rule with $f(a)= \frac{1}{a}$, a solo author can get a total of $k$ units of credit. However, adding $a-1$ spurious authors can fetch them a total of $\frac{ka}{H_a}$ credit, which grows in an unbounded fashion as $a$ increases.

The inequality~\eqref{Eqnfundamental} also
forms a natural boundary: Since a smaller per-author cost permits more
submissions, lowering $f(a)$ makes the rule more generous toward
collaborations---precisely the second principle we set out to achieve. The two
principles therefore push $f$ in opposite directions and meet at equality:
$f(a) = \frac{f(1)}{H_a}$ is the unique cost that is at once manipulation-resistant and
maximally generous to collaborations. 

We accordingly
take the coauthor-dependent component of the cost to be $\frac{1}{H_a}$, and set
$f(1) = 1$ without loss of generality, since scaling $f$ and $k$ together leaves
the rule unchanged.

\medskip 

In what follows, we extend our ideas so far on manipulation resistance to provide a more general result. Although this result is specified for $f(a) = \frac{1}{H_a}$, it continues to hold for the general Algorithm~\ref{alg:harmonic_quota} as well. Consider an unscrupulous researcher who wants to exploit the quota rule by padding author lists. They may begin with some papers already in hand, add spurious coauthors to reduce the quota cost of those papers, and then use the freed quota to submit additional papers in order to try and get more credit. To capture this scenario, we take the $m$ papers in the theorem below as the full set of papers the researcher could submit after carrying out such a manipulation. 
\begin{theorem}
Consider the credit model~\cite{hodge1981publication, hagen2008harmonic, hagen2010harmonic, hagen2013harmonic} where the r\textsl{a}nk-$i$ \textsl{a}uthor of \textsl{a}ny p\textsl{a}per with $a$ \textsl{a}uthors
receives credit $c(i,a)=\dfrac{1}{i\,H_a}$. Now consider \textsl{a} rese\textsl{a}rcher who h\textsl{a}s $m$ p\textsl{a}pers,
where p\textsl{a}per $\ell \in \{1,\ldots,m\}$ h\textsl{a}s $a_\ell$ \textsl{a}uthors \textsl{a}nd the rese\textsl{a}rcher holds \textsl{a}uthorship r\textsl{a}nk $i_\ell \in \{1,\ldots,a_\ell\}$. Suppose the rese\textsl{a}rcher p\textsl{a}ds more \textsl{a}uthors \textsl{a}nd repl\textsl{a}ces e\textsl{a}ch $a_\ell$ by some $a_\ell'\ge a_\ell$ while the rese\textsl{a}rcher rem\textsl{a}ins in r\textsl{a}nk
$i_\ell$.  Suppose submitting fr\textsl{a}ction\textsl{a}l p\textsl{a}pers m\textsl{a}y be permitted, \textsl{a}nd submitting \textsl{a} fr\textsl{a}ction $\phi_\ell\in[0,1]$ of p\textsl{a}per $\ell$ costs
$\phi_\ell f(\cdot)$ \textsl{a}nd yields credit $\phi_\ell c(i_\ell,\cdot)$. For \textsl{a}ny \textsl{a}uthorship sizes
$b=(b_1,\dots,b_m)$ let $C(b)$ denote the m\textsl{a}ximum credit th\textsl{a}t c\textsl{a}n be e\textsl{a}rned by the rese\textsl{a}rcher:
\[
C(b)=\!\!\max_{\phi\in[0,1]^m}\sum_{\ell\in \{1,\ldots,m\}}\phi_\ell\,c(i_\ell,b_\ell)
       \ \ \text{s.t.}\ \ \sum_{\ell\in \{1,\ldots,m\}} \phi_\ell f(b_\ell)\le k,
\]
Let $C^{\mathrm{int}}(b)$ be the s\textsl{a}me m\textsl{a}ximum when submissions c\textsl{a}nnot be fr\textsl{a}ction\textsl{a}l, th\textsl{a}t is, $\phi\in\{0,1\}^m$. 

In this setting, consider the quot\textsl{a} rule where e\textsl{a}ch \textsl{a}uthor h\textsl{a}s \textsl{a} budget $k$ \textsl{a}nd \textsl{a}ny p\textsl{a}per with $a$ \textsl{a}uthors costs e\textsl{a}ch of its \textsl{a}uthors $f(a)=\frac{1}{H_a}$. Then we h\textsl{a}ve:
\begin{enumerate}
\item[\textup{(\textsl{a})}] \emph{(Rese\textsl{a}rcher is le\textsl{a}d \textsl{a}uthor.)} If $i_\ell=1$ for \textsl{a}ll $\ell$, then for every
p\textsl{a}dding \textsl{a}nd every submission profile, the rese\textsl{a}rcher's tot\textsl{a}l credit is \textsl{a}t most $k$. Since $k$ is the credit \textsl{a}
le\textsl{a}d \textsl{a}uthor e\textsl{a}rns from single-\textsl{a}uthored submissions, p\textsl{a}dding c\textsl{a}nnot incre\textsl{a}se \textsl{a} le\textsl{a}d
\textsl{a}uthor's credit \textsl{a}bove the solo-\textsl{a}uthorship c\textsl{a}p.
\item[\textup{(b)}] \emph{(Fr\textsl{a}ction\textsl{a}l submissions.)} In \textsl{a} setting th\textsl{a}t \textsl{a}llows fr\textsl{a}ction\textsl{a}l submissions, for \textsl{a}ny \textsl{a}rbitr\textsl{a}ry \textsl{a}uthorship positions of the rese\textsl{a}rcher, p\textsl{a}dding c\textsl{a}nnot incre\textsl{a}se tot\textsl{a}l credit:
\begin{align*} C(a') \leq C(a).\end{align*}
\item[\textup{(c)}] \emph{(Integer submissions.)} When submissions must be integr\textsl{a}l, for \textsl{a}ny \textsl{a}rbitr\textsl{a}ry \textsl{a}uthorship positions of the rese\textsl{a}rcher, p\textsl{a}dding c\textsl{a}n yield \textsl{a}t most one \textsl{a}ddition\textsl{a}l p\textsl{a}per's worth of credit:
\[
  C^{\mathrm{int}}(a')
  \;\le\;C^{\mathrm{int}}(a)+\max_{\ell \in \{1,\ldots,m\}}c(i_\ell,a_\ell)
  \;\le\;C^{\mathrm{int}}(a)+1.
\]
\end{enumerate}
\label{ThmManipulationResistance}
\end{theorem}

The proof of this claim is provided in the appendix. In the next section we add a second, irreducible term that
lifts $f(a)$ above the boundary of~\eqref{Eqnfundamental}.


\subsection{An irreducible personal claim} 
A submission consumes scarce journal or conference resources on reviewers and editors as well as backend infrastructure.  Thus, when an author makes a submission, the author is using up the valuable resources of the journal, conference or the funder for evaluations. Some part of this is irreducibly on each author: a listed author still contributes to congestion in the journal's or conference's pipeline, has
their name attached to the request for reviewing attention, and receives some non-transferable benefit from being associated with the submission. It is this part -- which does not vanish even for very large author lists -- that has been the focus of past approaches with fixed quotas. Then there is another part which is shared across authors -- one submission with $a$ authors does not consume the same resources as $a$ submissions by one author each.

We therefore decompose the per-author claim into two components.  A cost
$\beta \geq 0$ is an irreducible personal claim, independent of the number of coauthors. This component is analogous to current practices of counting each submission as a unit cost on each author and bounding the total number of submissions any author can make. The remaining is a collective claim, shared across all authors, as investigated in the previous section. We thus write $f$ as a decomposition:
\[
f(a)=\beta+ \frac{1}{H_a}.
\]
It is important to note that the results of Theorem~\ref{ThmManipulationResistance}
continue to hold for this form. Intuitively, padding more authors does not reduce the cost $\beta$, thereby making padding disadvantageous and strengthening manipulation resistance.

\subsection{Computing the constants}
We thus have the form of our costs associated with submissions: 
\[
f(a)=\beta+\frac{1}{H_a},
\qquad \text{where~} 
H_a=\sum_{j=1}^a \frac{1}{j}.
\]
And each author has a total quota of $k$. We now derive the value of the constants $\beta$ and $k$. 

In order to do so, first consider the setting where large labs make many submissions. Consider a stylization of this setting where there are infinite authors in each submission. We focus on an author who is common to these submissions. The question is: how many submissions can such an author make? The organizer needs to provide such a value, which we will denote as $N_{\infty}$. Recall that if an author submits papers with $A$ authors each, then the number of papers they can submit is $\frac{k}{f(A)}$. With $A=\infty$, we then want
\[ 
k = N_{\infty} \lim_{a \rightarrow \infty} f(a)  = N_{\infty} \beta 
\]
since $ \lim_{a \rightarrow \infty} \frac{1}{H_a} = 0$. 
We thus set $\beta = \frac{k}{N_{\infty}}$. 

Next, let us choose the value of $k$. For this, consider single author submissions. These are the submissions which consume most evaluation resources per author. We thus ask organizers to specify another parameter, $N_1$. Consider an author who only makes single-authored submissions. Then $N_1$ is the number of such submissions that this author can make. Then we must have $k = N_1 f(1) = N_1 (\beta+1)$.

Solving the two equations relating $k$ and $\beta$ above, we get $\beta = \frac{N_1 }{N_{\infty}- N_1 }$ and $k = \frac{N_1 N_{\infty}}{N_{\infty}- N_1 }$.

\subsection{Putting it all together} 
Each submission with $a$ authors costs $f(a) = \frac{N_1 }{N_{\infty}- N_1 } + \frac{1}{H_a}$, and the total cost for an author should not exceed $k = \frac{N_1 N_{\infty}}{N_{\infty}- N_1 }$. We will rescale these values for better interpretablity: each submission with $a$ authors costs $f(a) = \frac{N_1 }{N_{\infty}} + \left( 1 -  \frac{N_1 }{N_{\infty}}\right) \frac{1}{H_a}$ to each author, and the total cost for any author should not exceed $k = N_1$. This results in the rule presented in Algorithm~\ref{alg:harmonic_quota}.  Note that setting $N_1 = N_{\infty}$ reduces this rule to the fixed rule. 

\section{Derivation of Generalized Harmonic Submission Quota (Algorithm~\ref{alg:generalized_harmonic_quota})}\label{Sec:General_derivation}

As discussed earlier, the parameter \(N_{\infty}\) is an asymptotic upper
bound, attained only for unrealistically large author lists. We now consider a setting where the organizers wish to instead specify a budget for  
a finite sized author list $A$. 

\subsection{Setting $N_\infty$} 
Our approach will be to calibrate $N_{\infty}$ based on the provided value of $N_A$, and call upon the Harmonic Quota Rule from Algorithm~\ref{alg:harmonic_quota}. 
With $f(a) \;=\; \frac{N_1}{N_{\infty}} \;+\; \left(1 - \frac{N_1}{N_{\infty}}\right)\frac{1}{H_a}$ and an initial quota of $N_1$, as specified in Algorithm~\ref{alg:harmonic_quota}, the number of submissions an author in a collaboration of size $A$ can make is $\frac{N_1}{\frac{N_1}{N_{\infty}} \;+\; \left(1 - \frac{N_1}{N_{\infty}}\right)\frac{1}{H_A}}$. In order to set $N_A$ as this value, some simple algebra yields that $N_{\infty}$ should be set as
\[
N_A
=
\frac{N_1}{
\frac{N_1}{N_{\infty}}
+
\left(1-\frac{N_1}{N_{\infty}}\right)\frac{1}{H_{A}}
},
\]
which then yields
\begin{align}
\label{EqNinfty}
N_{\infty}
=
\frac{ N_A (H_A-1)}{ H_{A}-\frac{N_A}{N_1}}. 
\end{align}

\subsection{The if condition}
For the value of $N_\infty$ in~\eqref{EqNinfty} to be valid, we need the denominator to be positive. 
When $H_{A,1} = H_A > \frac{N_A}{N_1}$, we simply call upon Algorithm~\ref{alg:harmonic_quota}, and set $N_\infty$ as the (finite) value obtained from~\eqref{EqNinfty}. The choice of $p=1$ reduces $H_{a,p}$ to the harmonic numbers $H_a$ used in Algorithm~\ref{alg:harmonic_quota}. This is the `if' condition in Algorithm~\ref{alg:generalized_harmonic_quota}.

\subsection{The else condition}
The case where the denominator of~\eqref{EqNinfty} is zero, that is $H_A = \frac{N_A}{N_1}$, is the boundary case between the if and else conditions (and can be handled by either of these conditions). In this case we have $N_\infty = \infty$. Algorithm~\ref{alg:harmonic_quota} then reduces to $f(a) = \frac{1}{H_a}$. 

Now consider $H_A < \frac{N_A}{N_1}$. Since $N_\infty$ in~\eqref{EqNinfty} will be invalid, we replace the harmonic number $H_A$ with its generalized version $H_{A,p}$. We then choose the largest $p \in [0,1]$ that meets $H_{A,p} \geq \frac{N_A}{N_1}$. Such a $p \in [0,1]$ exists and is unique because $\frac{N_A}{N_1} \in [1,A]$; $H_{A,p}$ is continuous and strictly decreases with an increase in $p$ (noting $A\geq 2$); and $H_{A,0}= A$ and $H_{A,1} < \frac{N_A}{N_1}$. At this chosen value of $p$, we have 
$H_{A,p} = \frac{N_A}{N_1}$, thereby yielding $N_\infty = \infty$. This is the choice made in the else condition of Algorithm~\ref{alg:generalized_harmonic_quota}.

\subsection{Putting it all together}
Now given the choices of $N_\infty$ and $p$ made above, and with the choice of $N_1$ specified by the organizer, we apply Algorithm~\ref{alg:harmonic_quota} but using the $p^{th}$ order harmonic numbers.

\section{Conclusions and future work}\label{Sec:Practical}
We derived a quota rule and its generalization from first principles. If organizers impose any such quota, authors may have to choose between their papers. There is some evidence that authors can do a good job of this~\cite{su2025icml}, although different coauthors may not always agree on the relative merits of their jointly authored papers~\cite{rastogi2022authors}. We now discuss some more considerations when deploying this rule in practice, alongside avenues for future work.  

First, in venues with rolling submissions like journals, one could consider schemes that provide some quota back to authors of accepted submissions. For authors who are also contributing to the evaluations as reviewers, editors etc., one may start with a higher quota for their service. Incorporating these additional aspects in a more principled manner is left as future work. 

Second, we make no assumptions about the relation between ordering of authors and their relative contributions. For instance, some submissions may choose to have the lead contributor as first author whereas others may order authors in alphabetical or random order. If these orderings follow certain standards, one may in the future explore submission rules that do not treat all authors of a submission on the same footing. 

Third, there are many other types of potential strategic gaming that should be investigated. For instance, we would also like to prevent exploitation of quotas via exchange of authorship. Specifically, suppose two authors add each other to their respective submissions, then their total quota should not increase. Suppose their papers originally had $a_1$ and $a_2$ authors. The two authors together originally used $f(a_1) + f(a_2)$ of their combined budget. By adding each other, they now together use up $2f(a_1+1) + 2f(a_2+1)$ of their combined budget. One can verify that for every $a_1, a_2$, under our Harmonic Quota Rule, we have $2f(a_1+1) + 2f(a_2+1) > f(a_1) + f(a_2)$, thereby dis-incentivizing this type of manipulation. 

Fourth, many evaluation systems are moving to explicitly randomize their decisions~\cite{heyard2022rethinking,goldberg2026principled}. In such settings, when a proposal is rejected partly because of randomization rather than perceived quality, organizers must decide whether the corresponding quota should be consumed, restored, or treated differently.

Finally, our proposed rule can work in conjunction with various other approaches to address increasing submissions such as AI reviewing~\cite[Part 1]{shah2025aiPeerReview}, submission fees~\cite{ijcai2026cfp}, proportional reciprocal reviewing~\cite{arr2025incentives}, restrictions on resubmissions~\cite{erc2026resubmission}.

\section*{Acknowledgments}
This work was funded by grants   NSF 1942124 and ONR N000142512346.

\bibliographystyle{alpha}
\bibliography{bibtex}

\appendix
~\\\noindent{\bf \Large Appendix: Proof of Theorem~\ref{ThmManipulationResistance}}\\

\emph{Part (\textsl{a}).} Let $i_\ell=1$ for all $\ell$ as in the premise of the claim that the researcher is the lead author on all their papers. Let $b_\ell\in\{a_\ell,a_\ell'\}$ be any
(padded or unpadded) sizes. Then we have $c(1,b_\ell)=f(b_\ell)$, so for every
budget-feasible $\phi$ (fractional or integer) it must be that
\[
  \sum_{\ell \in \{1,\ldots,m\}} \phi_\ell\,c(1,b_\ell)=\sum_{\ell\in \{1,\ldots,m\}} \phi_\ell f(b_\ell)\le k .
\]
The value $k$ is attained by
single-authored submissions, since $f(1)=c(1,1)=1$. Hence no padding lets a lead
author exceed the solo cap of $k$.\\

\emph{Part (b).} We will perform a reparameterization that will allow us to compare the optimization programs under the original author list $a$  and the padded one $a'$. For every $\ell \in \{1,\ldots,m\}$, define 
$x_\ell=\phi_\ell f(\cdot)\in[0,f(\cdot)]$. Then the credit that the researcher gets from paper
$\ell$ is $\phi_\ell c(i_\ell,\cdot)=x_\ell/i_\ell$. This yields
\[
  C(a)=\max_{x \in \mathbb{R}^m} \sum_{\ell\in \{1,\ldots,m\}} \frac{x_\ell}{i_\ell}
  \ \text{~~~~~s.t.}\ \textstyle\sum_{\ell\in \{1,\ldots,m\}} x_\ell\le k, \text{~~~and~~~}\ 0\le x_\ell\le f(a_\ell)~\forall \ell \in \{1,\ldots,m\},
\]
\[
  C(a')=\max_{x \in \mathbb{R}^m} \sum_{\ell\in \{1,\ldots,m\}} \frac{x_\ell}{i_\ell}
  \ \text{~~~~~s.t.}\ \textstyle\sum_{\ell\in \{1,\ldots,m\}} x_\ell\le k, \text{~~~and~~~}\ 0\le x_\ell\le f(a_\ell')~\forall \ell \in \{1,\ldots,m\}.
\]
The objective is the same in both optimization programs. As for the constraints, we have $f(a_\ell')\le f(a_\ell)$, and hence the feasible region for $C(a')$ is contained
in that for $(a)$. Hence we have $C(a) \geq C(a')$.\\

\emph{Part (c).} Since integer profiles are feasible for the fractional program, we have
$C^{\mathrm{int}}(a') \le C(a')$. From part (b), we also have 
$C(a') \le C(a)$. It remains to bound
$C(a) -C^{\mathrm{int}}(a)$. 

Let us now look at the optimization program for $C(a)$. The program $\max_x\sum_\ell x_\ell/i_\ell$ is a linear program in $m$ variables, subject to one budget constraint $\sum_\ell x_\ell\le k$ and $m$ box constraints
$0\le x_\ell\le f(a_\ell)$. A bounded linear program attains an optimum at a vertex, where $m$ linearly independent constraints will be tight. Since at most one of
these is the budget constraint, at least $m-1$ tight constraints are box bounds which set its coordinate to $0$ or $f(a_\ell)$. Hence at most one coordinate lies
strictly inside its box, i.e., at most one coordinate is fractional. So under the optimal solution, a set (say, $S$) of papers is submitted fully and (at most) one paper (say, $t$) at fraction $\phi_t\in[0,1)$. Submitting exactly $S$ is integer-feasible and has credit
\[
  C(a) -\phi_t\,c(i_t,a_t)\;\ge\;C(a) -\max_\ell c(i_\ell,a_\ell),
\]
so $C^{\mathrm{int}}(a) \ge C(a)-\max_\ell c(i_\ell,a_\ell)$.
Chaining the inequalities, we get the claimed bound
\[
  C^{\mathrm{int}}(a') \le C(a')\le C(a) 
  \le C^{\mathrm{int}}(a)+\max_{\ell \in \{1,\ldots,m\}} c(i_\ell,a_\ell),
\]
and $\max_\ell c(i_\ell,a_\ell)\le c(1,1)=1$. This yields the claimed result.

\end{document}